\documentclass[preprint,preprintnumbers,amsmath,amssymb,12pt,floatfix,epsfig,nofootinbib,superscriptaddress]{revtex4}

\def\be{\begin{equation}}
\def\ee{\end{equation}}
\def\Be{\begin{eqnarray}}
\def\Ee{\end{eqnarray}}
\def\ba{\begin{array}}
\def\ea{\end{array}}

\usepackage{amsmath} 
\usepackage{bm}      
\usepackage[usenames, dvipsnames]{color}
\usepackage{graphicx}
\usepackage{amssymb}
\usepackage[normalem]{ulem}
\renewcommand\sout{\bgroup\color[rgb]{1,0,0} \ULdepth=-.5ex \ULset}
\begin{document}

\title{KDAR neutrino scattering for $^{12}$C target via charged current and muon angular distribution}

\author{Chaeyun Lee \footnote{\textrm{e-mail:} cylee31005@gmail.com}}
\address{Department of Physics and OMEG Institute, Soongsil University, Seoul 06978, Korea}

\author{Kyungsik Kim \footnote{\textrm{e-mail:} kyungsik@kau.ac.kr}}
\address{School of Liberal Arts and Science, Korea Aerospace University, Koyang 10540, Korea}

\author{Myung-Ki Cheoun \footnote{\textrm{e-mail:} cheoun@ssu.ac.kr (Corresponding Author)}}
\address{Department of Physics and OMEG Institute, Soongsil University, Seoul 06978, Korea}

\author{Eunja Ha \footnote{\textrm{e-mail:}  ejaha@hanyang.ac.kr}}
\address{Department of Physics and Research Institute for Natural Science, Hanyang University, Seoul, 04763, Korea}

\author{Tatsushi Shima \footnote{\textrm{e-mail:} shima@rcnp.osaka-u.ac.jp}}
\address{Research Center for Nuclear Physics, Osaka University, Osaka 567-0047, Japan}

\author{Toshitaka Kajino \footnote{\textrm{e-mail:} kajino@buaa.edu.cn}}
\address{School of Physics and International Research Center for Big-Bang
 Cosmology and Element Genesis, Beihang University, Beijing 100083, China,
 The University of Tokyo, 7-3-1 Hongo, Bunkyo-ku, Tokyo 113-0033, Japan,
 National Astronomical Observatory of Japan 2-21-1 Osawa, Mitaka, Tokyo 181-8588, Japan}

\date{\today}
\begin{abstract}

We investigate the charged current (CC) scattering of muon neutrinos ($\nu_{\mu}$) off $^{12}$C using an incident neutrino energy of $E_{\nu_{\mu}} =$ 236 MeV, as produced in kaon decay-at-rest (KDAR). In this energy regime, both quasielastic (QE) scattering and inelastic processes occurring below the QE region contribute simultaneously. To account for these contributions, we calculate the inelastic scattering cross section using the quasiparticle random phase approximation (QRPA) and combine it with the QE scattering cross section obtained via the distorted wave Born approximation (DWBA) within the framework of relativistic mean field (RMF) theory. The results are compared with MiniBooNE experimental data. Additionally, as the outgoing muon in KDAR 
$\nu_{\mu}$ CC scattering may exhibit angular dependence, we analyze the differential cross section as a function of the scattering angle. These findings may provide valuable input for the calibration of ongoing KDAR neutrino cross-section measurements.

\end{abstract}

\maketitle

\section{Introduction}

Neutrino ($\nu$) (antineutrino (${\bar \nu}$))-induced reactions and the $\nu$-scattering on complex nuclei play a crucial role in advancing our understanding of nuclear structure as probed by the weak interaction \cite{Freed2003,Oconn1972,Suzuk2006,Suhonen11,Kosmas11,Kosmas11-2,Volpe00}. Additionally, these reactions provide insights into key neutrino properties relevant to neutrino physics, such as the neutrino mass hierarchy, matter effects in neutrino oscillations, and neutrino self-interactions. Moreover, they offer constraints on the composition of stellar matter through detailed analyses of nuclear abundances in core-collapse supernova (SN) explosions \cite{Woosl1990,Yoshi2008,Moto2019,Ko2022}.

Recently, significant attention has been directed toward the $\nu$-process \cite{Woosl1990,Yoshi2008,Heger2005,Moto2019,Ko2022,Suzuk2009} in medium and heavy nuclei as well as light nuclei. Despite the relatively small weak interaction cross sections, the high neutrino flux emitted in astrophysical environments is expected to facilitate the production of specific nuclei that are otherwise blocked by $\beta$-decay due to the presence of surrounding stable seed nuclei. Consequently, cross sections for neutrino and antineutrino interactions with nuclei ($\nu ({\bar \nu})–A$) serve as essential input data for network calculations estimating the nuclear abundances, particularly in the weak rapid process \cite{Wanaj2006,Ch10,Ch11-2}.

Beyond nucleosynthesis, a detailed understanding of neutrino-nucleus scattering cross sections is also vital for the interpretation of data from recent accelerator- and reactor-based neutrino experiments \cite{Alva2018}.

In accelerator-based neutrino experiments, most neutrinos are produced via decay-in-flight (DIF) and three-body beta decay of pions and/or kaons generated in proton-nucleus interactions. As a result, the neutrino energy spectrum is broad, posing a significant challenge in experimental analyses, as it necessitates the reconstruction of the incident neutrino energy using a specific model. Beyond the complexities associated with energy reconstruction, a precise determination of the nuclear response functions to electroweak interactions is crucial for assessing nuclear model dependencies in the interpretation of experimental observables. This is particularly important given the limited understanding of weak-interaction-driven nuclear structure and the inherently small cross sections involved.



Kaon-decay-at-rest (KDAR) neutrinos provide an optimal framework for studying neutrino-nucleus interactions, free from the complications associated with the broad energy distribution of neutrinos originating from meson decay-in-flight (DIF). Beyond their significance in sterile neutrino searches \cite{Atha1997,JSNS1,JSNS2,Aqui2018,JSNS2-2024}, the monoenergetic 236-MeV $\nu_{\mu}$ beam serves as a precise tool for the unambiguous calibration of cross sections and the determination of weak-interaction parameters within nuclei. Consequently, KDAR $\nu_{\mu}$ neutrinos hold considerable potential for significantly reducing both experimental and theoretical uncertainties, thereby mitigating ambiguities in neutrino interaction studies with unprecedented precision.


KDAR neutrinos also play a crucial role in advancing our understanding of astrophysical neutrinos. High-energy proton accelerators primarily produce neutrinos via meson DIF, resulting in GeV-scale neutrino energies that are too high to effectively study supernova neutrinos, which are expected to have energies in the tens of MeV range. Furthermore, neutrino interactions with nuclei in this intermediate energy region remain largely unexplored, with only limited experimental data available from laboratory-produced neutrinos, such as those from LSND \cite{Atha1997} and KARMEN \cite{Bodm1994,Masc1998,Armb1998}.


Due to their relatively low energy, 236 MeV KDAR $\nu_{\mu}$ neutrinos offer a valuable opportunity to investigate neutrino-nucleus interactions relevant to core-collapse supernovae and their associated neutrino processes \cite{Ko2022,Moto2019}, as well as sterile neutrino searches \cite{Aqui2001,Aqui2018}. KDAR $\nu_{\mu}$ interactions produce cross sections with energy transfers up to $E_{\nu_{\mu}} - m_{\mu} = $130 MeV, encompassing a range of nuclear responses consistent with the energy domain of supernova neutrinos. In particular, the forward scattering of 236 MeV $\nu_{\mu}$ with low-momentum transfer provides crucial insights into low-energy nuclear excitations, which play a significant role in astrophysical neutrino interactions. These studies complement experiments conducted with $\nu_{\mu}$ decay-at-rest (DAR) neutrinos at lower energies, around $E_{\nu_{\mu}} \sim$ 30 MeV \cite{Hara2013,Elni2013,Aqui2021,Scho2020,Akim2017,Dist2003}.


In this energy region, neutrino-nucleus ($\nu–A$) scattering involves contributions from both the one-step quasielastic (QE) scattering process and the two-step inelastic scattering process \cite{Kim2009}. In the two-step process, the incident neutrino initially excites the target nucleus, which subsequently decays into other nuclear states through the incoherent emission of secondary particles. This excitation occurs via various multipole transitions, including superallowed Fermi 
($J^{\pi} = 0^{+}$), allowed Gamow-Teller ($J^{\pi} = 1^{+}$), spin dipole ($J^{\pi}$ = ${0}^{-}$, ${1}^{-}$, ${2}^{-}$), and higher-order multipole transitions. The dominant contributions to the two-step process arise from discrete nuclear states and giant resonances (GR) of the compound nucleus, with typical excitation energies on the order of tens of MeV.

In contrast, the one-step process occurs when an individual nucleon within the target nucleus is directly ejected due to sufficient momentum transfer, with minimal excitation of the residual nucleus. This mechanism is the primary reaction channel in the QE peak region, where the incident neutrino scatters quasi-freely off nucleons. If the outgoing particles in the two-step process are nucleons, the two reaction mechanisms become experimentally indistinguishable, as both result in identical final states.

For example, $^{12}$C($\nu , {\nu}^{'}$)$^{12}$C${}^{*}$
 $\rightarrow $ $^{11}$B + p (or $^{11}$C+n) reaction via neutral current (NC) could not be distinguished from
$^{12}$C$(\nu, {\nu}^{'} p) ^{11}$B (or $^{12}$C$(\nu, {\nu}^{'} n) ^{11}$C). Similarly, $^{12}$C$({\nu}_e, e^{-}) {}^{12}$N$^{*} \rightarrow ^{11}$C + p and $^{12}$C (${\bar \nu}_e, e^{+}) ^{12}$B$^{*} \rightarrow ^{11}$B + n reactions through charged current (CC) also
could not be differentiated, respectively, from the QE scattering $^{12}$C$({\nu}_e, e^{-} p)$ $^{11}$C and
$^{12}$C$({\bar \nu}_e$, $e^{+} n$) $^{11}$ B. This one-step process may affect the nucleosynthesis by supernovae neutrino which has considered mostly the two-step process \cite{Kim2009}.


Experimental studies of KDAR neutrinos have recently commenced, with several existing and planned experiments designed to exploit this unique neutrino source. One prominent KDAR $\nu_{\mu}$ source is the NuMI beamline absorber at Fermilab, where multiple neutrino detectors in its vicinity are capable of detecting these neutrinos. Notably, the MiniBooNE experiment, a Cherenkov- and scintillation-based mineral oil detector located approximately 85 m from the NuMI beamline absorber, has successfully identified 236-MeV KDAR $\nu_{\mu}$ events using muon energy reconstruction and timing information \cite{Aqui2018}. This led to the first measurement of monoenergetic KDAR $\nu_{\mu}$ interactions on $^{12}$C, reported by the MiniBooNE collaboration through shape-only differential cross sections \cite{Aqui2018}.

Further advancements in KDAR neutrino studies have been made by the J-PARC Sterile Neutrino Search at the J-PARC Spallation Neutron Source (JSNS$^2$) project, which utilizes a 3-GeV pulsed proton beam as a high-intensity KDAR neutrino source \cite{JSNS1,JSNS2}. The first experimental results from JSNS$^2$ are now available in Ref. \cite{JSNS2-2024}, providing cross-section measurements as a function of missing energy and/or visible energy.


In this study, we investigate the differential cross sections for charged current quasielastic (CCQE) and inelastic scattering of 236 MeV $\nu_{\mu}$ KDAR neutrinos off $^{12}$C. Notably, the relevant energy transfer for these interactions lies in the low-energy region of the genuine QE domain. Consequently, nuclear effects and low-energy excitations are expected to play a significant role, necessitating a detailed microscopic modeling of these neutrino interactions with careful treatment of nuclear structure effects \cite{Niko2021}.

To describe low-energy neutrino interactions with nuclei via inelastic and QE scattering, we employ the quasiparticle random phase approximation (QRPA) for inelastic scattering and utilize relativistic mean field (RMF) models within the distorted wave Born approximation (DWBA) framework for QE scattering. Additionally, we note that the kinematic phase space accessible in KDAR neutrino scattering off $^{12}$C spans 50 $< q <$ 400 MeV and 17 $< \omega <$ 130 MeV, where the kinematical conditions for the two-step inelastic process described by QRPA and the one-step QE process modeled by RPA overlap, depending on the scattering angle \cite{Niko2021}.


Furthermore, the outgoing muon in KDAR neutrino scattering may exhibit angular dependence due to the large momentum transfer, offering valuable insights into weak nuclear structure. To account for this effect, we evaluate the angular distribution of the outgoing lepton, explicitly incorporating the Coulomb interaction between the emitted lepton and the residual nucleus.

\section{Theoretical Framework}
By using the weak current operator, which is composed of
longitudinal (${\cal {\hat L}}_J$), Coulomb (${\cal {\hat C}}_J$), electric ($ {\cal
{\hat T}}_J^{el} $) and magnetic operators (${\cal {\hat T}}_J^{mag} $) detailed at Ref.\cite{Ch09-2}, we calculate the differential cross section for $\nu ({\bar \nu})$-$^{12}$C
reactions as follows \cite{Ch09-1,Don79,Wal75}
\Be & & ({{d \sigma_{\nu}} \over {d \Omega }  })_{(\nu / {\bar
\nu})} = { { G_F^2 \epsilon k } \over {\pi ~ (2 J_i + 1 ) }}~
\bigl[ ~ {\mathop\Sigma_{J = 0}} (
 1+ {\vec \nu} \cdot {\vec \beta }){| <  J_f || {\cal {\hat C}}_J || J_i > | }^2
 \\ \nonumber & & + (
 1 - {\vec \nu} \cdot {\vec \beta } + 2({\hat \nu} \cdot {\hat q} )
 ({\hat q} \cdot {\vec \beta}  ))
  {| <  J_f || {\cal {\hat L}}_J ||
J_i > | }^2  - \\ \nonumber & & {\hat q} \cdot ({\hat \nu}+ {\vec
\beta} )  { 2 Re < J_f || {\cal {\hat L}}_J  || J_i>
{< J_f|| {\cal {\hat C}}_J || J_i >}^*  } \\
\nonumber & &  + {\mathop\Sigma_{J = 1}} ( 1 - ({\hat \nu} \cdot
{\hat q} )({\hat q} \cdot {\vec \beta}  ) ) ( {| <  J_f || {\cal
{\hat T}}_J^{el}  || J_i > | }^2 + {| <  J_f || {\cal {\hat
T}}_J^{mag} || J_i > | }^2
) \\
\nonumber & &  \pm {\mathop\Sigma_{J = 1}} {\hat q} \cdot ({\hat
\nu} - {\vec \beta} )  2 Re [ <  J_f || {\cal {\hat T}}_J^{mag} ||
J_i > {<  J_f || {\cal {\hat T}}_J^{el} || J_i > }^* ]\bigr]~, 
\label{eq:Xsec}\Ee
where $(\pm)$ stems from the different helicities of $\nu ({\bar \nu})$, respectively.
${\vec \nu}$ and $ {\vec k}$ are incident and final lepton
3-momenta, ${\vec q} = {\vec \nu} - {\vec k}$ and ${\vec \beta} =
{\vec k} / \epsilon $ with the final lepton's energy $\epsilon$. The nuclear matrix elements of the weak interaction Hamiltonian can be expanded in
terms of multipole operators by using two basic operators
\be { M}_J^{M_J} ( q {\bf x})  =  j_J ( q { x}) Y_J^{M_J} (
\Omega_x )~, ~  {\bf M}_{J L}^{M_J} ( q {\bf x})  = j_J ( q { x})
{\bf Y}_{J L 1}^{M_J} ( \Omega_x )~,\label{eq:sepa} \ee
where vectorial spherical harmonic ${\bf Y}_{J L 1}^{M_J} (
\Omega_x )$ is expressed in term of spherical harmonic $Y_{L}^{m}
( \Omega_x )$, {\it i.e.} ${\bf Y}_{J L 1}^{M_J} ( \Omega_x ) =
 {\mathop\Sigma_{m \lambda}} < L m 1
\lambda \vert ( L 1 ) J M_J
> Y_{L}^{m} ( \Omega_x ) {\bf e}_{\lambda} $. Then
we rewrite any one-body transition matrix element ${\cal {\hat
O}}_{JM;TM_T}^{ (1) } ( q {\bf x}) $ in terms of the 4 different
transition operator (Coulomb, longitudinal, electric and magnetic)
as follows
\Be {\cal { {\hat C}}}_{JM;T M_T} (q {\bf x}) & = &  [ F_1^{(T)}
M_J^{M_J} ( q {\bf x} ) - i { q \over M} [ { F_A^{(T)}
{{\Omega}}_J^{M_J} ( q {\bf x} ) + {{ F_A - \omega F_P^{(T)}}
\over 2} {\Sigma^{''}}_{J}^{M_J} ( q {\bf x}) } ]] I_T^{M_T},
\\ \nonumber
~{\cal {{\hat L}}}_{JM;T M_T} (q {\bf x}) & = & [{- {\omega  }
\over { q } } F_1^{(T)} M_J^{M_J} ( q {\bf x} ) + i ( F_A^{(T)} -
{{q^2  }\over {2 M_N } } F_P^{(T)} ) {\Sigma^{''}}_{J}^{M_{J}} ( q
{\bf x} ) ]I_T^{M_T}~,\\
\nonumber {\cal { {\hat T}}}_{JM;T M_T}^{el} (q {\bf x}) & = & [{
q \over M} [ { F_1^{(T)} {{\Delta}^{'}}_J^{M_J} ( q {\bf x} ) + {1
\over 2} \mu^{(T) } {\Sigma}_{J}^{M_J} ( q {\bf x}) }] + i
F_A^{(T)} {\Sigma^{'}}_{J}^{M_{J}} ( q {\bf x} ) ] I_T^{M_T}~,
\\ \nonumber
{\cal { {\hat T}}}_{JM;T M_T}^{mag} (q {\bf x}) & = & -i { q \over
M} [[ { F_1^{(T)} {{\Delta}}_J^{M_J} ( q {\bf x} ) - {1\over 2}
\mu^{(T) } {\Sigma^{'}}_{J}^{M_J} ( q {\bf x}) }] +   F_A^{(T)}
{\Sigma}_{J}^{M_{J}} ( q {\bf x} ) ] I_T^{M_T}~,
\label{eq:multi} \Ee
where the 7 relevant single particle operators ($ M_J^{M_J}
,{{\Omega}}_J^{M_J}, {\Sigma^{''}}_{J}^{M_J}, {{\Delta}}_J^{M_J},
{\Sigma}_{J}^{M_J}, {\Sigma^{'}}_{J}^{M_J}, {{\Delta}^{'}}_J^{M_J}
$) are detailed at Refs.\cite{Con72,Don79}. The superscript $T ( =
0,1) $ means isoscalar and isovector. $I_T^{M_T}$ stands for the
isospin dependence \cite{Don79}. Single nucleon form factors
$F_X^{(T)} (Q^2) $ with $T = 0,1$ and $X = 1,2,S,A,P,T$ are Dirac
($X=1$), Pauli ($X=2$), scalar, axial, pseudo-scalar, and tensor
form factors, respectively. Detailed form factors are referred
from Refs. \cite{Con72,Don79}, and we take $F_S$ and $F_T$ to
be zero because of current conservation and no existence of second class current, respectively. Here we note that these operators include the GT and Fermi transitions including the forbidden parts as well as other spin dipole resonance transitions. The multipole transitions are calculated by the QRPA framework \cite{Ch93}. The results  for $\nu_e$-$^{12}$C scattering in Refs. \cite{Ch10,Ch09-2} turned out to agree with the shell model (SM)calculation \cite{Suzuk2006,Suzuk2009}.

For CC reactions we multiplied Cabbibo angle $cos ^2 \theta_c$ and
considered the Coulomb distortion of outgoing leptons in a
residual nucleus \cite{Suzuk2006,Suzuk2009,Ring08}. By following the
prescriptions adopted in Refs.\cite{Ring08,Kolbe03-a}, we choose an energy
point in which both Fermi function and the effective momentum
approach (EMA) approaches predict same values. Then we use
the Fermi function below the energy and the EMA above the energy \cite{Co05-2,Co06,Angel98}. But for the QE scattering region, we exploit the exact treatment of the Coulomb distortion performed in Refs. \cite{Kim97,Kim2011}.

Here we make a note on the meaning of the $1^+$ transition exploited in the neutrino cross section in Eq.(1), which is used differently from the allowed GT($1^+$) transition for the $\beta$-decay or charge exchange reactions. For neutrino-induced reaction cross sections, we include the forbidden $1^+$ transitions as well as the allowed GT transition, while only the allowed GT transition is usually considered at the GT strength distributions. Hereafter, we denote the corresponding $1^+$ transition for neutrino reaction as $1^+$ transition to distinguish the allowed GT transition. Of course, the $1^+$ transition is reduced to the allowed GT transition at the threshold limit.
It means that the Ikeda sum rule (ISR) useful for the GT transition may not hold in the $\nu$ and ${\bar \nu}$ cross sections. For the neutrino-induced reactions, we did not use the quenching factor for any multipole transitions because the ISR for the GT transition is more or less satisfied if we include the high-lying GT excitations beyond one- and two-nucleon separation energy \cite{Yako2003}, and we do not know the quenching factors for other multipole transitions.

For the $\nu_{\mu}$-$^{12}$C scattering in the QE region we exploit the DWBA approach based on a relativistic mean field (RMF) model and quark meson coupling (QMC) model, whose formula are briefly introduced in the following in order to compare to the QRPA case. 

We choose that the target nucleus is seated at the origin of the coordinate system.
The four-momenta of the participating bodies are denoted as $p_i^{\mu}=(E_i, {\bf p}_i)$, $p_f^{\mu}=(E_f, {\bf p}_f)$, $p_A^{\mu}=(E_A, {\bf p}_A)$, $p_{A-1}^{\mu}=(E_{A-1}, {\bf p}_{A-1})$, and $p^{\mu}=(E_N, {\bf p})$ for the incident neutrino, outgoing muon, target nucleus, the residual nucleus, and the knocked-out nucleon, respectively.
Within the laboratory frame, the inclusive cross section in the CC reaction, which detects only outgoing lepton, is given by the contraction between lepton and hadron tensors:
\begin{eqnarray}
{\frac {d\sigma} {dT_N}} &=& 4\pi^2{\frac {M_N M_{A-1}} {(2\pi)^3
M_A}} \int \sin \theta_l d\theta_l \int \sin \theta_N d\theta_N p
f^{-1}_{rec} \sigma^{W^{\pm}}_M [ v_L R_L  + v_T R_T + h v_T' R_T' ],
\label{cs}
\end{eqnarray}
where $M_N$ is the nucleon mass in free space, $\theta_l$ denotes the scattering angle of the lepton, $\theta_N$ is the polar angle of knocked-out nucleons, $T_N$ is the kinetic energy of the knocked-out nucleon, and $h=-1$ $(h=+1)$ corresponds to the intrinsic helicity of the incident neutrino (antineutrino).
The $R_L, R_T$ and $R^{'}_T$ are longitudinal, transverse, and transverse interference response functions, respectively.
Detailed forms for the kinematical coefficients $v$ and the corresponding response functions $R$ are given in Ref. \cite{Kim2023}.
The squared four-momentum transfer is given by $Q^2=q^2 - \omega^2=-q^2_{\mu}$.
For the CC reaction, the kinematic factor $\sigma^{W^\pm}_M$ is defined by

\begin{equation}
\sigma^{W^\pm}_M = \sqrt{1 - {\frac {M^2_l} {E_f}}} \left ( {\frac
{G_F \cos (\theta_C) E_f M_W^2} {2\pi (Q^2 + M^2_W)}} \right )^2,
\end{equation}
where $M_W$ is the rest mass of $W$-boson, and $M_l$ is the mass of an outgoing lepton.
$\theta_C$ represents the Cabibbo angle given by $\cos^2 {\theta_C} \simeq 0.9749$.
$G_F$ denotes the Fermi constant.
The recoil factor $f_{rec}$ is written as
\begin{equation}
f_{rec} = {\frac {E_{A-1}} {M_A}} \left | 1 + {\frac {E_N}
{E_{A-1}}} \left [ 1 - {\frac {{\bf q} \cdot {\bf p}} {p^2}}
\right ] \right |. 
\end{equation}
Here we shortly introduce the RMF model adopted in the present work. The quantum hadrodynamics (QHD), as a representative RMF nuclear model, has been established by Walecka within a relativistic framework
for describing nuclear many-body system, where the point like nucleons interact through the exchange of scalar ($\sigma$)
and vector ($\omega$) meson \cite{33}. 

Guichon \cite{38} propose the quark-meson coupling (QMC) model, in which the properties of nuclear
matter can be self-consistently calculated by the coupling of  meson fields to the quarks within the nucleons rather than to
the nucleons themselves \cite{39}. 
 
These RMF models, denoted as RMF and QMC, were successfully applied by present authors to the exclusive $(e,e' p)$ \cite{42} and inclusive $(e,e')$ \cite{43} reactions for electron-nucleus QE scattering and also shown to be in good agreement with the neutrino QE scattering data \cite{Kim2022,Kim2023}. The detail explanations of each model are done in Refs. \cite{42,Kim2023}. Recently the QMC models are extended by including the isovector-scalar $\delta$ meson giving rise to the charge symmetry breaking effects and applied to the neutron star physics \cite{Miya2023,Miya2022}. In this work, we exploit the QRPA approach for inelastic scattering and the DWBA approach based on the RMF models for describing the KDAR neutrino QE scattering.

\section{Results and discussions}

\begin{figure}[ht]
  \includegraphics[width=13.5cm]{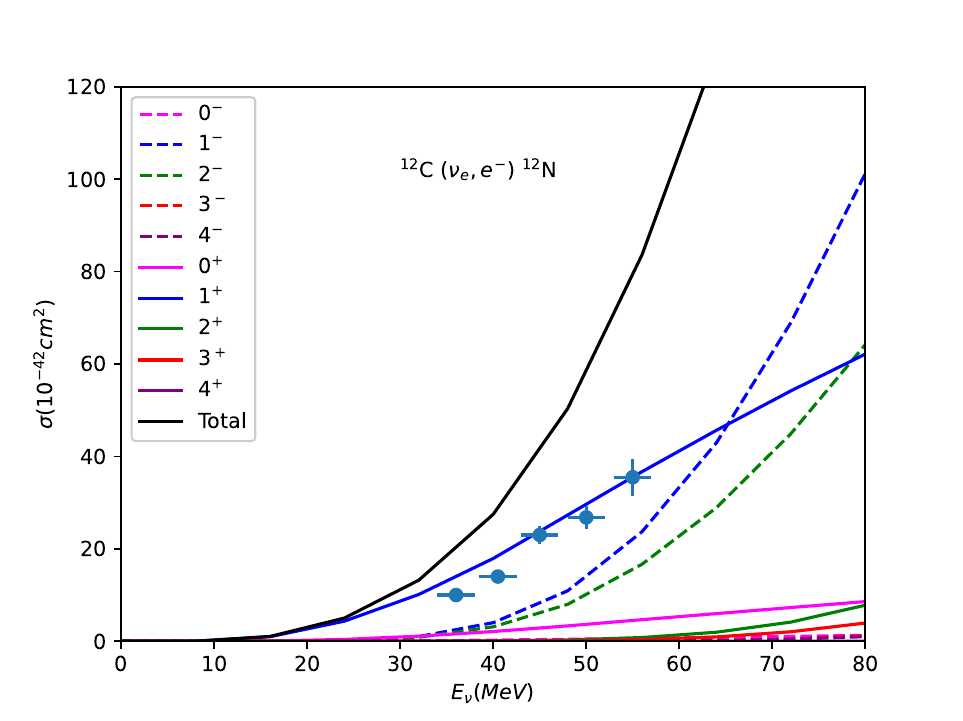}
\caption{(Color online) Neutrino-induced cross section via CC for $^{12}$C, $^{12}$C$(\nu_e, e^-)^{12}$N$^*$. Total sum from $J^{\pi} = 0^{\pm}$ up to $J^{\pi} = 4^{\pm}$ state transitions and each contribution to the total transition are presented. They are compared to the LSND experimental
 data, $^{12}$C$(\nu_e, e^-)^{12}$N$_{g.s. (1^+)}$ taken from Ref. \cite{LSND}.}
\label{fig:1}
\end{figure}

\subsection{Cross sections of KDAR neutrino scattering off $^{12}$C in terms of muon kinetic energy $T_{\mu}$}  

\begin{figure}[ht]
  \includegraphics[width=8.0cm]{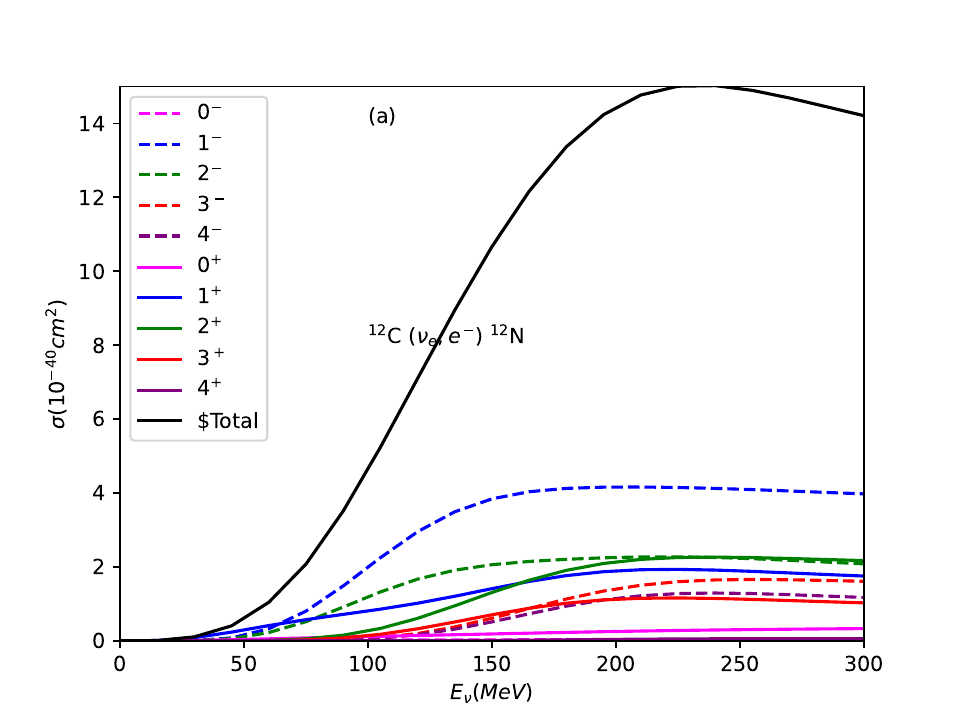}  
  \includegraphics[width=8.0cm]{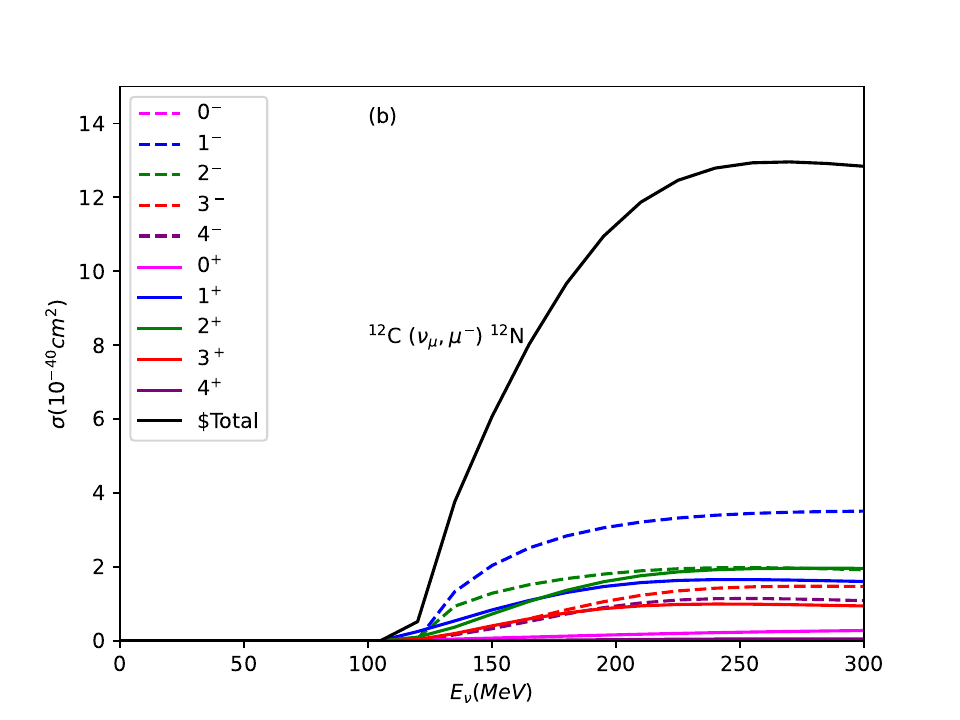}
\caption{(Color online) Neutrino-induced cross section via CC (a) for $^{12}$C, $^{12}$C$(\nu_e, e^-)^{12}$N$^*$ and (b) $^{12}$C$(\nu_{\mu}, {\mu}^-)^{12}$N$^*$ upto $E_{\nu}$= 300 MeV}
\label{fig:2}
\end{figure}

Figure \ref{fig:1} illustrates total cross section of $^{12}$C($\nu_e , e^-) ^{12}$N, which were already provided in previous papers \cite{Ch09-1,Ch09-2} and shown to be consistent with the results by the shell model calculation in Ref.\cite{Suzuk2006}. Experimental data was given for a specific channel,  $^{12}$C$(\nu_e, e^-)^{12}$N$_{g.s. (1^+)}$ \cite{LSND}, and explained well by the $1^+$ transition obtained by the QRPA calculation. One point to notice is that the contribution by the $1^+$ transition measured by detecting $^{12}$N$^*$ \cite{LSND} dominates the cross section up to $E_{\nu} \simeq 60$ MeV. After the region, $1^{-}$ spin-dipole (SD) transition becomes dominant with the $2^{-}$ SD transition contribution.

Before the discussion of the cross section by KDAR $\nu_{\mu}$, in Fig.\ref{fig:2}, we extend the calculation up to $E_{\nu} = 300$ MeV within the QRPA scheme to study the neutrino flavour dependence in the total cross section data. We could not find any significant dependence on neutrino flavour. But interestingly, the main transition is changed to the spin dipole $1^-$ transition above $E_{\nu}$ = 60 MeV, similarly to the results in Fig. \ref{fig:1}, irrespective of the neutrino flavour, and the muon production threshold energy explicitly appeared.  

Other spin dipole transition $2^-$ as well as the $1^+$ and $2^+$ transitions contribute to some extent to total cross section. Above $E_{\nu_e} \sim$ 150 MeV ($E_{\nu_{\mu }} \sim$ 250 MeV), most of the transitions are saturated. It means that the nuclear excitations above the energy region are already exhausted, and the QE scattering becomes dominant above the energy region. This QE contribution is detailed by the RMF and QMC models later on.

\begin{figure}[ht]
  \includegraphics[width=8cm]{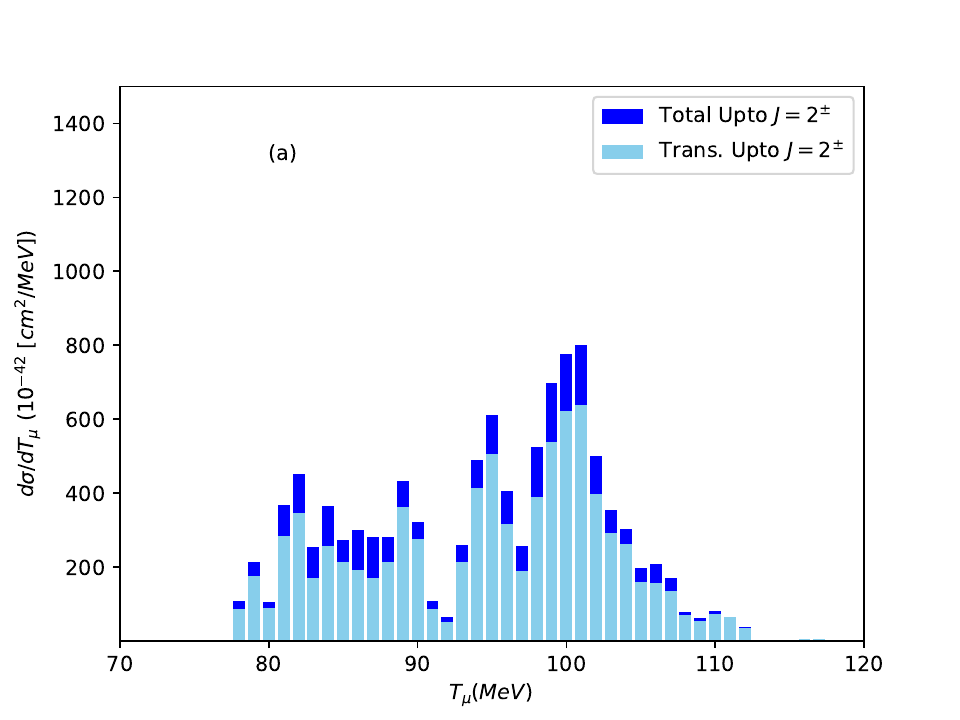}
  \includegraphics[width=8cm]{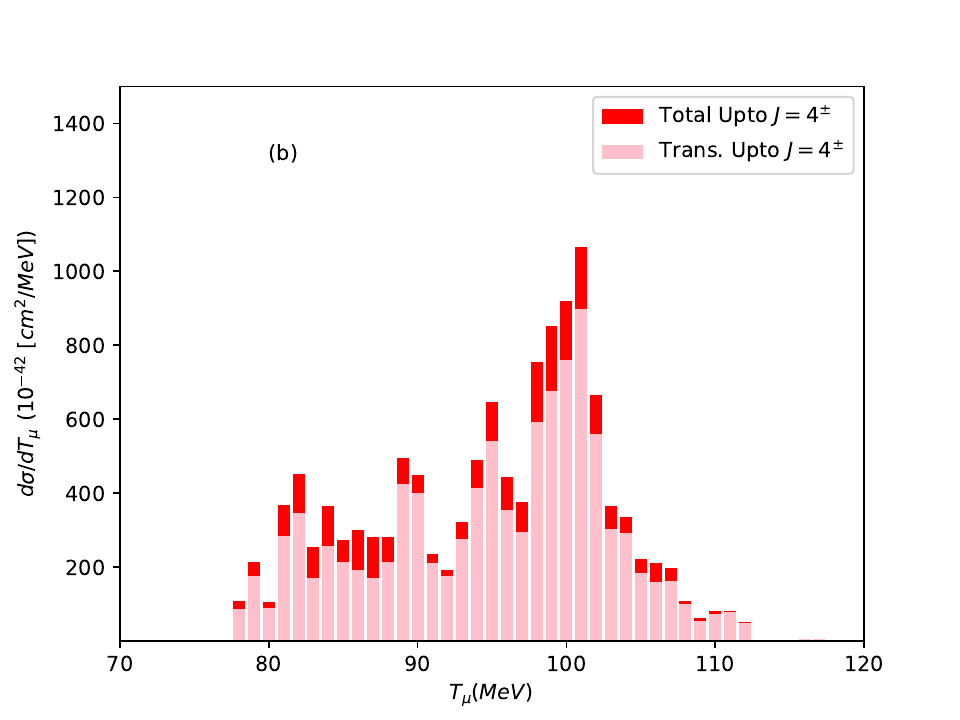}
\caption{(Color online) Neutrino cross section via CC for $^{12}$C by KDAR neutrino in term of the muon kinetic energy $T_{\mu}$ from $J = 0^{\pm}$ up to (a) $J_{max}$ = $2^{\pm}$ and (b) $4^{\pm}$, respectively. The contribution from the electro-magnetic transverse part is represented in sky-blue (a) and pink color (b) in each panel.}
\label{fig:3}
\end{figure}

Hereafter, we focus on the differential cross section by KDAR, whose energy is fixed as $E_{\nu_{\mu}}=236$ MeV. Figure \ref{fig:3} (a) and (b) present the differential cross section from $J = 0^{\pm}$ up to, respectively, $J^{\pi} = 2^{\pm}$ and $J^{\pi} = 4^{\pm}$. The cross sections are distributed mostly in the region, 78 MeV $<$ $T_{\mu}$ $<$ 113 MeV, and the most pronounced peaks appear around $T_{\mu} =$100 MeV region.

Here we shortly introduce the kinematics used in this work to grasp the concept of physical variables of KDAR $\nu_{\mu}$ scattering. If we start from the missing energy concept, $E_m = \omega - \sum {T_p}$, usually used in the electron scattering, $^{12}$C$(e,e'p)$, with the energy transfer $\omega$ and the outgoing kinetic energy of protons and muons, $T_p$ and $T_{\mu}$, then the missing energy in the neutrino scattering is given by the kinetic energies, {\it i.e.}   
\begin{equation}
E_m = 129.2 MeV - ( T_{\mu} + \sum {T_p})~,~ \omega = 129.2 MeV - T_{\mu}~.
\end{equation}
Therefore, the missing energy and energy transfer can be obtained by measuring the kinetic energy of outgoing particles.

In this work, we take kinetic energy of emitted proton from $^{12}$N$^*$ produced in the inelastic scattering as the nuclear excitation energy, {\it i.e.} $E^* = \sum {T_p}$. Then we obtain $E_{\nu_\mu} = m_{\mu} + T_{\mu} + E^* + Q$ with the $Q$ value (= 16.32 MeV) between $^{12}$C and $^{12}$N.  The cross section in Fig.\ref{fig:3} given by the muon kinetic energy $T_{\mu}$ appears around $ T_{\mu} = 78 \sim 113$ MeV. It means that the excitation energy spectra $E^* = 0 \sim 35$ MeV region of $^{12}$N$^{*}$ are taken into account in the compound nucleus $^{12}$N produced by KDAR $\nu_{\mu}$. We also note that about 20\% increase obtained by $J^{\pi} > 2^{\pm}$ transitions was found in the cross section results. 

We divide the contribution of the response functions, longitudinal, Coulomb and electro-magnetic transverse parts in Eq. (1). The transverse part contribute largely about 80 \% to the the cross section rather than the Coulomb and longitudinal part in the KDAR neutrino kinematics.

\begin{figure}[ht]
  \includegraphics[width=13.5cm]{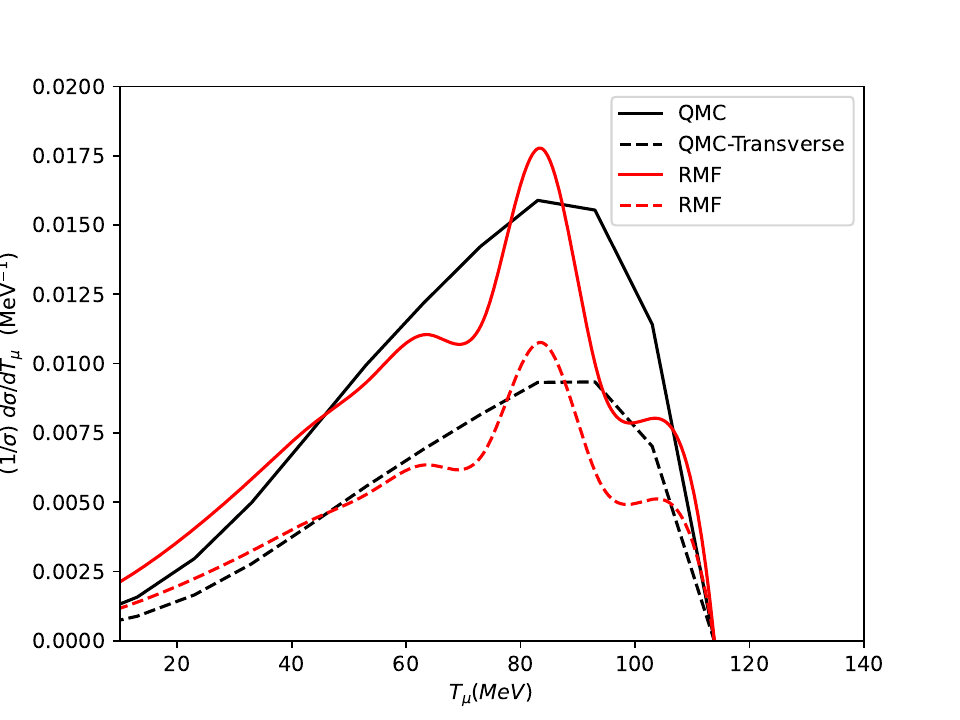}
\caption{(Color online) Neutrino cross sections via CC for $^{12}$C by KDAR neutrino by the DWBA based on RMF (red) and QMC (black) model with the contribution by the transverse component. Solid and dotted lines stand for total and transverse cross sections, respectively.}
\label{fig:4}
\end{figure}

\begin{figure}[ht]
  \includegraphics[width=13.5cm]{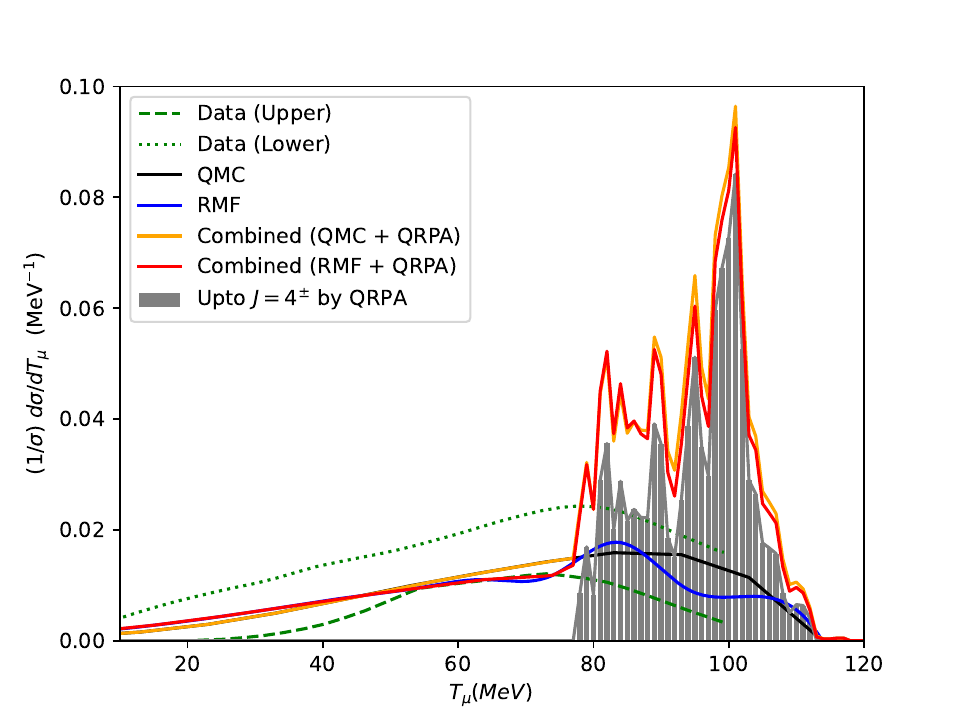}
\caption{(Color online) Neutrino cross sections via CC for $^{12}$C by KDAR neutrino in the DWBA based on the QMC (black) \cite{Kim2023} and RMF (blue) model \cite{Wal75}, and QRPA (gray histograms). DWBA calculation corresponds to the one-step (curves) and the QRPA calculation corresponds to two-step (gray histograms) processes for the neutrino scattering.} The upper and lower green curves show the experimental data by the shape spectrum from MiniBooNE KDAR data \cite{Aqui2018}.
\label{fig:5}
\end{figure}

Figure \ref{fig:4} presents the results by DWBA calculation for one-step process, {\it i.e.} QE scattering. The DWBA results are obtained by two  different RMF models \cite{Kim2022}. One (red curve) is the simple RMF model adopted in Ref. \cite{33} and the other model (black curve) is the so called QMC model \cite{Kim2023} which included the density dependence by the $\sigma$ meson in the RMF model considering the quark condensation in vacuum, and has been successfully used for neutrino and electron scattering \cite{Kim2023}. As shown in Fig.\ref{fig:4}, the transverse part in Eq. (\ref{cs}) contributes about 50 \% to the cross section, contrary to the 80 \% contribution in the inelastic scattering shown in Fig.\ref{fig:3}.

Finally, in Fig. \ref{fig:5}, we present total cross section by the QRPA and DWBA approach. 
The cross section denoted as bar graphs in $T_{\mu}=$ 78 $\sim$ 113 MeV (corresponding to $E^*$ = 0 $\sim$ 35 MeV) are obtained by the QRPA approach for the inelastic scattering taking into account of the excitations in $^{12}$N. The cross sections from $T_{\mu}$ = 10 MeV up to 113 MeV are obtained by the DWBA considering QE scattering. Here the experimental data are given by the upper and lower curves taken from Ref. \cite{Aqui2018}. 

In brief, the significant peaks in the KDAR $\nu_{\mu}$-$^{12}$C scattering data from MiniBooNE and JSNS$^2$ experiments \cite{Aqui2018,JSNS2-2024} are explained mostly by the inelastic scattering, while the lower $T_{\mu}$ part comes from the QE scattering. 
In the overlapped $T_{\mu}$ region, the QE scattering contributes maximally 25 \% and minimally 13 \% to the sum of cross section. Therefore, in this energy region both contributions should be taken into account for the $\nu$-reaction as argued in our previous paper \cite{Kim2009}.

\subsection{Angular distribution of outgoing muon in KDAR scattering}

\begin{figure}[ht]
  \includegraphics[width=12.5cm]{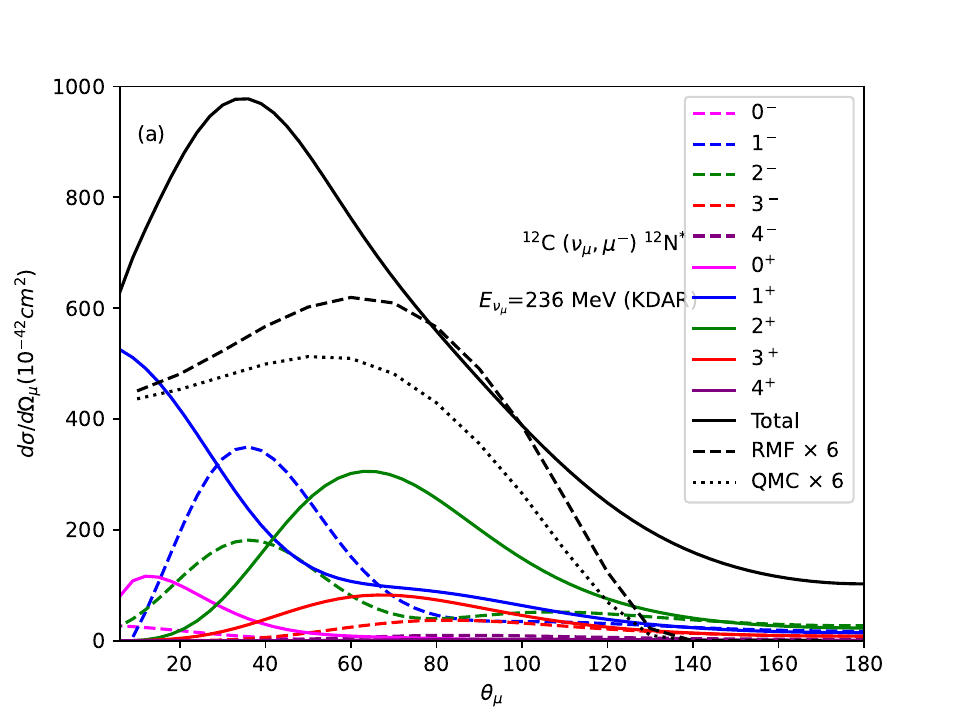}
  \includegraphics[width=12.5cm]{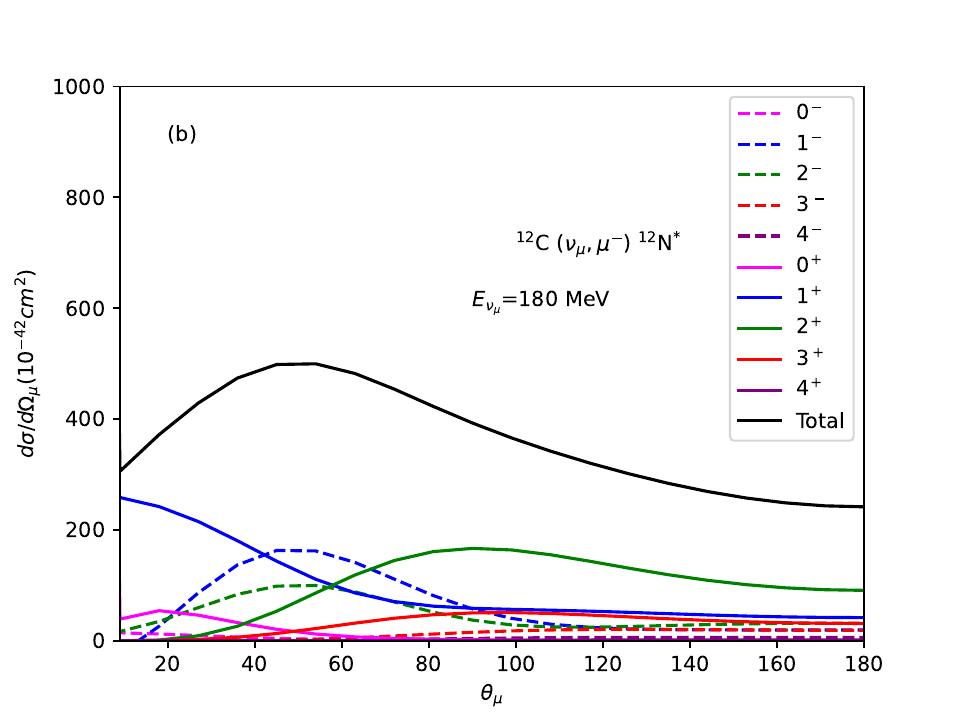}
\caption{(Color online) Muon angular distribution cross section via CC (a) for $^{12}$C for KDAR neutrino ($E_{\nu_\mu}=$ 236 MeV) for each $J^{\pi}$ = $0^{\pm}, 1^{\pm},2^{\pm},3^{\pm},4^{\pm}$ transition, respectively. The contribution from DWBA based on the RMF and QMC model are also presented in black dashed and dotted lines, respectively, enlarged by 6 times, in the panel (a).} In order to understand the energy dependence of the angular distribution, the results for $E_{\nu_{\mu}}$= 180 MeV are also presented in the panel (b) without the DWBA results.
\label{fig:6}
\end{figure}

Figure \ref{fig:6} displays the angular distribution of outgoing lepton, respectively, for $E_{\nu_{\mu}}$ = 236 MeV (a) and 180 MeV (b). One can see forward peak around $\theta_{\mu}$ = 40$^o$, whose peak values (positions) increase (shift to forward) with the increase of the incident energy. Main contributions around this peak angle region stem mainly from the spin dipole transition $1^-$ and the quadrupole transition $2^-$. But, interestingly, the $1^+$ transition show a prominent forward peak, while the $2^+$ transition provide a relatively backward angle distribution. Namely, if we lower the neutrino energy as $E_{\nu_\mu} =$ 180 MeV in the panel (b), the peak position is shifted a bit backward to $\theta_{\mu} =$ 50$^0$ region mainly due to the shift by $1^-$, $2^-$, and $2^+$ transitions. The QE scattering also shows a bit forward peak. But the contribution turns out to be about 10 \%, similarly to the results in Fig. \ref{fig:5} given by $T_{\mu}$ variable.

\begin{figure}[ht]
  \includegraphics[width=8.0cm]{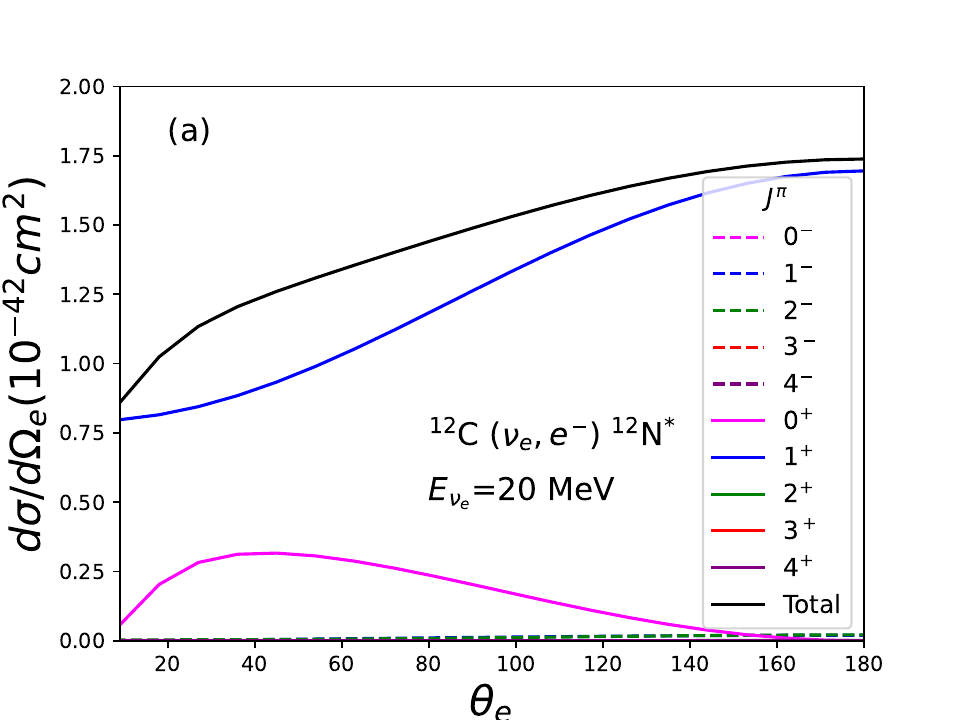}
  \includegraphics[width=8.0cm]{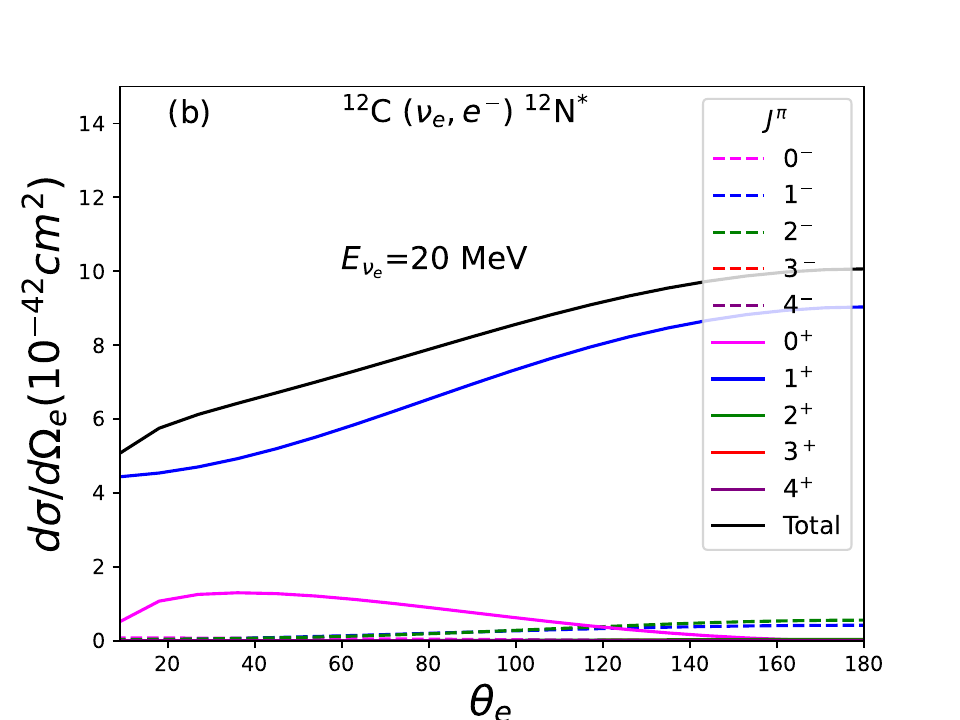}
  \includegraphics[width=8.0cm]{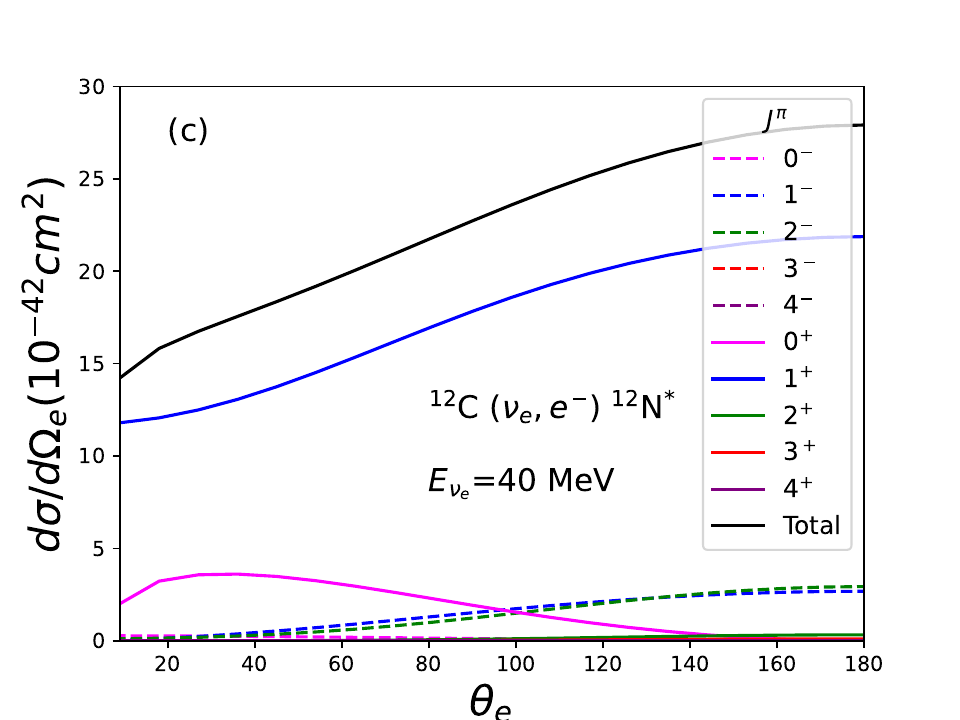}
  \includegraphics[width=8.0cm]{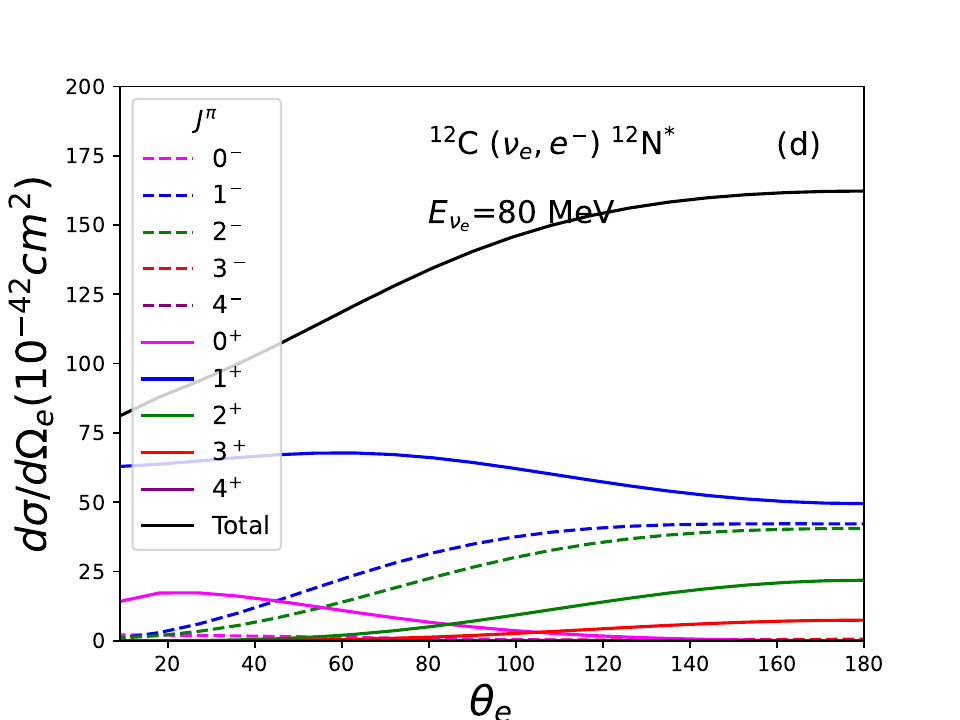}
\caption{(Color online) Electron angular distribution cross section for $\nu_e$-$^{12}$C reaction for each $J^{\pi}$ = $0^{\pm},1^{\pm},2^{\pm},3^{\pm}$, and $4^{\pm}$ transition, respectively.}
\label{fig:7}
\end{figure}

However, the situation becomes quite different for the $\nu_e$ case. The results for $\nu_e$ in Fig.\ref{fig:7} show the increase at backward direction, almost irrespective of the incident energy. This is totally different from the trend of muon behaviour. The difference comes from the lepton mass difference. The heavy lepton (muon) moves to the forward direction, but the lighter electron moves backward in the scattering off the $^{12}$C target. Important point for $\nu_e$ case is that the main transition contributing to the angular distribution comes from the $1^+$ transition in the energy region $E_{\nu_e} <$ 80 MeV. But with the increase of incident neutrino energy, the $1^+$ transition shifts forward direction as shown in the panel (d), while other multipole transitions except $0^+$ transition moves still backward direction with the enlarged contribution. Consequently, the angular distribution shows more backward direction. This trend is quite different from those by the $\nu_{\mu}$ case in Fig. \ref{fig:6}. But it is interesting that the $0^+$ transition shows forward direction independent of neutrino flavour. 

Another point to deserve notice is that the results for the $\nu_e$ case are dominated by the two $1^+$ and $0^+$ transitions. Above $E_{\nu_e} \simeq$ 40 MeV, the spin dipole dominance ($1^-$ and $2^-$) appears with backward direction contrary to the $\nu_{\mu}$ case in Fig. \ref{fig:6}. The same spin dipole transition shows different angular distribution tendency depending on the neutrino flavour. 

Recent calculation by Fermi gas model with RPA corrections presented backward peaks for both $\nu_e$ and $\nu_{\mu}$ cases \cite{Akbar2017}. But the calculation did not include explicitly the multipole transitions. The calculation by CRPA show backward peaks for $\nu_e$ scattering for 50 $\sim$ 100 MeV.  This behaviour is compatible with the present calculation. But with the increase of the $E_{\nu_e}$ about 300 $\sim$ 500 MeV, whose energy region is mostly QE scattering region, forward peaks are shown to be prominent for $\nu_e$-$^{16}$O scattering \cite{Kolbe03-a}.

\section{Summary and Conclusion}

First, we calculated KDAR neutrino scattering off a $^{12}$C target using the relativistic mean field (RMF) and quark-meson coupling (QMC) models within the distorted wave Born approximation (DWBA) for the quasielastic (QE) region, and the quasiparticle random phase approximation (QRPA) for the region below the QE threshold. The results for the QE region successfully reproduced the MiniBooNE data up to approximately $T_{\mu} =$80 MeV, as described by the RMF approach \cite{Kim2022,Kim2023}.

Second, in the region where $T_{\mu} > $80 MeV, inelastic scattering contributions become dominant. In addition to the QE scattering contribution, we included the inelastic scattering contribution based on the QRPA approach, considering the excitation of the compound nucleus resulting from the inelastic scattering of the incident KDAR neutrino. The QRPA calculation adequately explained the high peaks observed in the $T_{\mu}$= 78 $\sim$ 113 MeV range. This QRPA approach has also been successfully applied to describe neutrino-induced cross sections for various target nuclei in neutrino-processes \cite{Ko2022, Moto2019}.

Third, with the expectation of more data from low-energy neutrino beam facilities, such as the JSNS$^{2}$ experiment, we investigated the angular distribution of the outgoing lepton. The results reveal interesting characteristics. The outgoing muon from KDAR neutrino scattering predominantly exhibits forward scattering, primarily due to the $1^{\pm}$ and $2^{\pm}$ transitions, although each multipole transition has distinct angular dependencies. As the incident energy decreases, the peak positions shift toward the backward direction. In contrast, electron neutrino scattering predominantly displays backward scattering patterns. The difference between forward and backward scattering cross sections is nearly a factor of two. This disparity arises from the mass difference between the outgoing leptons. The heavier muon is more likely to scatter in the forward direction, whereas the lighter electron primarily scatters in the backward direction. We note that the angular dependence of the outgoing leptons may influence the recoil of the target nuclei and could be observable in future experiments.

\section*{Acknowledgements}
This work was supported by the National Research Foundation of Korea (NRF) grant funded by the Korea government (Grant No. RS-2021-NR060129).

\end{document}